\documentclass[aps,pre,twocolumn,superscriptaddress,floatfix,showpacs]{revtex4}
\usepackage{graphicx}
\usepackage{bm}
\bibstyle{apsrev.bib}

\begin{document}
\title{Unbinding of mutually avoiding random walks and two dimensional quantum gravity}
\author{Enrico Carlon}
\affiliation{Interdisciplinary Research Institute c/o IEMN, Cit\'e
Scientifique BP 69, F-59652 Villeneuve d'Ascq, France}
\author{Marco Baiesi}
\affiliation{INFM Dipartimento di Fisica, via Marzolo 8, I-35131 Padova, Italy}

\date{\today}

\begin{abstract}
We analyze the unbinding transition for a two dimensional lattice polymer
in which the constituent strands are mutually avoiding random walks.  At low
temperatures the strands are bound and form a single self-avoiding walk.
We show that unbinding in this model is a strong first order transition.
The entropic exponents associated to denaturated loops and end-segments
distributions show sharp differences at the transition point and in
the high temperature phase. Their values can be deduced from some
exact arguments relying on a conformal mapping of copolymer networks
into a fluctuating geometry, i.e. in the presence of quantum gravity. An
excellent agreement between analytical and numerical estimates is observed
for all cases analized. 
\end{abstract}

\maketitle

\newcommand{\bc}{\begin{center}}
\newcommand{\ec}{\end{center}}
\newcommand{\be}{\begin{equation}}
\newcommand{\ee}{\end{equation}}
\newcommand{\ba}{\begin{array}}
\newcommand{\ea}{\end{array}}
\newcommand{\beqn}{\begin{eqnarray}}
\newcommand{\eeqn}{\end{eqnarray}}

\section{Introduction}

The unbinding transition from a low temperature double stranded polymer
to a high temperature single stranded phase has been the subject
of recent attention in the context of studies of DNA denaturation
\cite{cule97,kafr00,caus00,gare01,carl02,barb03}. Two main approaches
have been used to model this unbinding. The first one relies on
the use of directed polymers, where only the transversal coordinate
measuring the distance between homologous base pairs is considered
\cite{peyr89,cule97}. In a second approach, one considers the polymer as
being composed of an alternating sequence of double stranded segments and
denaturated loops \cite{pola66,kafr00}. The statistical weights assigned
to loops and segments can be estimated using concepts of homopolymers
and self-avoiding walks (SAW) statistics.

Traditionally, in the latter class of models, the statistical weight of
a loop was approximated as the number of configurations for a closed
SAW, neglecting any excluded volume interaction with the rest of the
chain \cite{fish66}. More recently, statistical mechanical ideas based
on the theory of polymer networks \cite{dupl86}, were used to take into
account the excluded volume effects in an approximated way \cite{kafr00}.
Although quite simple, this analysis, which was carried out analytically,
captures quantitatively very well the asymptotic form of the loop
partition function, as a series of numerical investigations on two and
three dimensional lattice models have shown \cite{carl02,baie02}. Despite
the general good agreement between analytical predictions and simulations,
quite small but systematic deviations between the two were found
\cite{baie02}.

The aim of this paper is to investigate further on these issues for other
types of models of polymer unbinding.  We focus on a two dimensional
lattice model for which we obtain a series of analytical predictions based
on exact results from conformal invariance \cite{dupl99}. In this model,
the two constituent strands are two random walks (RWs) with an attractive
interaction. While we relax the excluded volume interactions within each
strand, mutual avoidance, i.e. the non-overlapping condition between
the strands is preserved.  This makes the bound double stranded state
to behave as a SAW, assigning to it a quite different physics compared
to that of loops. We show that in this model unbinding is a very strong
first order transition.

\begin{figure}[b]
\includegraphics[height=3cm]{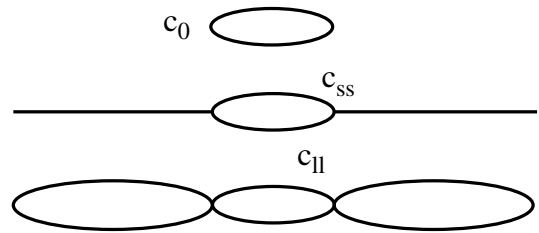}
\caption{Hierarchy of exponents describing the loop entropy for an
isolated loop ($c_0$), a loop embedded between two long segments ($c_{\rm
ss}$) and a loop embedded between two long loops ($c_{\rm ll}$).}
\label{FIG01}
\end{figure}

As is well-known from polymer physics, the partition function of a closed
SAW of total length $l$ assumes the following asymptotic form \cite{vand98}
\be
Z(l) \sim \mu^l l^{-c_0}
\label{ring}
\ee
where $\mu$ is a geometric factor and $c_0$ a universal exponent which
equals $c_0 \approx 1.76$ \cite{fish66} in three dimensions.  It has
been shown \cite{kafr00} that a loop attached to two long segments
or loops (see Fig.~\ref{FIG01}) has still a partition function of the
form of Eq. (\ref{ring}), but with different exponents.  For instance
in three dimensions one finds for a loop embedded between two long
segments \cite{kafr00} $c_{\rm ss} \approx 2.1$ and $c_{\rm ll}
\approx 2.2$ for a loop embedded between two long loops (here we used
the subscripts $s$ or $l$ to indicate neighboring segments or loops).
An increase of $c$ for embedded loops is caused by the tendency of the
loop to become more "localized" due to excluded volume interactions
with the rest of the chain. For $c > 2$ the transition becomes first
order \cite{pola66,fish66}.

\begin{figure}[t]
\includegraphics[height=6cm]{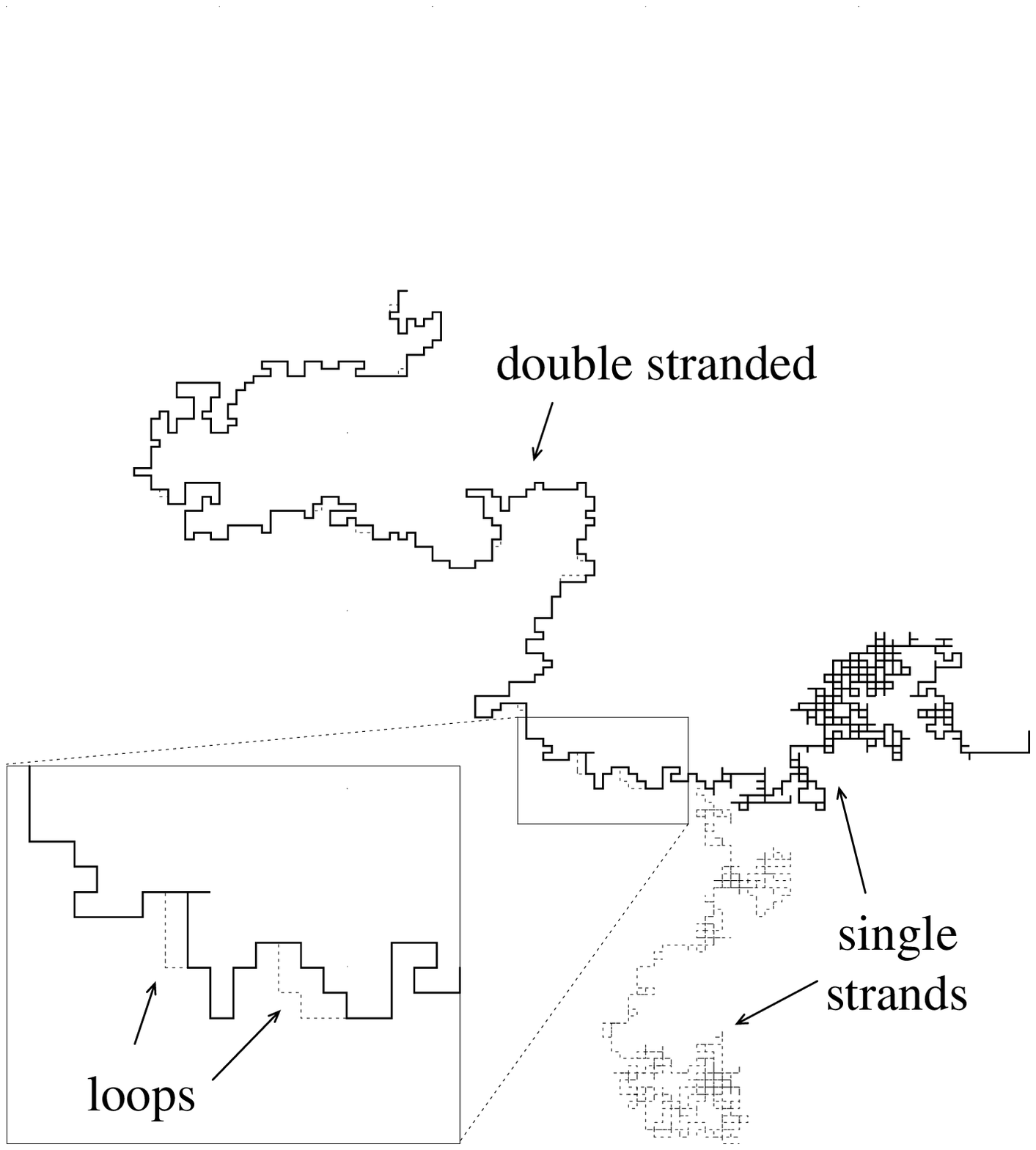}
\caption{Snapshot of a configuration for the model studied in this paper
generated by the Pruned Enriched Rosenbluth Method (PERM). The two
walks are indicated as solid and dashed lines. When they are
bound only the solid line is visible. Inset: Blow up of a small area of
double stranded phase with two embedded loops.
The loops are oriented, i.e. following the direction of one of the strands,
the other strands is always found on the same side.
}
\label{FIG02}
\end{figure}

In this paper we calculate analytically the values of the exponents
$c_{\rm ll}$ and $c_{\rm ss}$, as well as other type of entropic
exponents, for the model of mutually avoiding random walks. The
calculation relies on conformal mapping of copolymer networks into a
fluctuating geometry \cite{dupl99}, i.e. in the presence of quantum
gravity. This theory, which was recently applied to the calculation of
multifractal spectra of harmonic measures \cite{dupl03}, enables one to
obtain exact entropic exponents of networks composed of arbitrary mixtures
of random and self-avoiding walks \cite{dupl99}. We also performed a
series of Monte Carlo simulations in order to determine the numerical
values for the exponents. In all cases analyzed the numerical and
analytical results are in excellent agreement. This is due, as we
will discuss below, to the strong first order nature of the unbinding
transition, which makes the model an ideal testground where copolymer
network theories can be applied.

Part of the results presented here, have been discussed in concise form in
Ref.~\cite{baie02b}. In this paper we present the results of large scale
numerical calculations, which we extend to other quantities not considered
previously, and present a full account of the analytical results. The
present model has also been investigated by means a continuum approach
\cite{gare01}, in which mutual avoidance between the two strands has
been approximated by an effective long-range interaction, an approach
which predicts a first order unbinding transition.

\section{The model}
\label{sec:model}

We consider two random walks of length $N$ on a square lattice described
by the vectors $\vec{r}_1 (k)$ and $\vec{r}_2 (k)$ with $k=0,1 \ldots
N$. The walks have common origin $\vec{r}_1 (0) = \vec{r}_2 (0)$ and
are not allowed to overlap except at homologous sites, i.e. $\vec{r}_1
(i) = \vec{r}_2 (j)$ is possible only if $i=j$. Whenever such a contact
is realized the system gains an energy $\varepsilon = -1$.  At very low
temperatures the two walks are fully bound and form a self-avoiding walk,
since, as the walks are mutually avoiding, a bound site $\vec{r}_1 (i)
= \vec{r}_2 (i)$ cannot overlap any other sites on both strands. At
higher temperatures unbinding starts from the unconstrained edge $i=j=N$
and loops proliferate along the chain. Figure \ref{FIG02} shows a snapshot
of a configuration generated by the Pruned Enriched Rosenbluth Method
(PERM) \cite{gras97}, where the two strands are represented by solid and
dashed lines, respectively.  Notice that the double-stranded part has a
characteristic self-avoiding walk behavior and the unbound single strands
behave as mutually avoiding random walks.  The inset of Fig.~\ref{FIG02}
shows a blow-up of part of the double stranded chain with two short loops
of few lattice spacings of length. The model is constructed such that
following the two strands $\vec{r}_1 (k)$ and $\vec{r}_2 (k)$ from the
origin $k=0$ to $k=N$ one finds one of the two strands (say $\vec{r}_1
(k)$) when unbound always at the left side of the other strand, i.e. the
loops are oriented. This choice will allow us to restrict the type of
diagrams considered in the continuum polymer network description of
the model.

\begin{figure}[t]
\includegraphics[height=5cm]{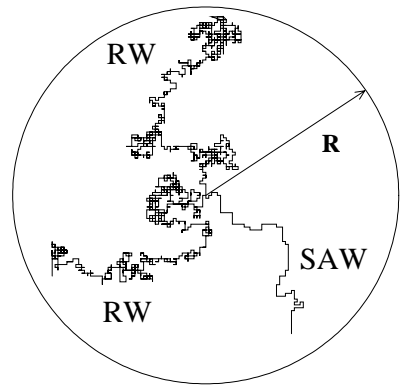}
\caption{
Example of a star copolymer formed by two random walks and a self-avoiding
walk, where all three of them are mutually avoiding and have average
size equal to $R$. The exact entropic exponent for any star copolymer
with an arbitrary number of random and self-avoiding walks can be
calculated thanks to a mapping onto a fluctuating geometry (see
Sec. \ref{sec:networks}).
}
\label{FIG03}
\end{figure}
\section{Conformal mapping onto two dimensional quantum gravity}
\label{sec:networks}

\subsection{Star copolymers}

Duplantier formulated an elegant theory \cite{dupl99} which allows one
to compute exact entropic exponents for star copolymers, an example
of which is shown in Fig.~\ref{FIG03}. A star copolymer is formed
by an arbitrary mixture of random and self-avoiding walks joined at a
common origin, which can be all mutually avoiding (as in the example of
Fig.~\ref{FIG03}) or partially transparent to each other.

One is typically interested in the grand-canonical partition function $Z_R
(S)$, for a star $S$ formed by $f_1$ self-avoiding walks (SAWs) and $f_2$
random walks (RWs), all with average size $R$ from the origin of the
star, $R$ denoting the end-to-end distance for each walk.  Having the
same size, SAWs and RWs are characterized by different lengths, scaling
as $\sim R^{1/\nu}$, with $\nu=3/4$ for a SAW and $\nu=1/2$ for a RW.
Fugacities per unit of step length are associated to each type of
walk. By tuning their values appropriately both walks become critical.
In this limit the partition function scales asymptotically for large $R$
as \cite{dupl99}
\be
Z (S) \sim R^{\eta(S) - f_1 \eta_2} ,
\label{star_disc}
\ee
where $\eta(S)$ is the scaling exponent associated to the singularity
at the center of the star and $\eta_2$ the entropic exponent associated
to an isolated SAW ($\eta_2 = -11/24$ in two dimensions, while the
corresponding exponent for RWs is zero). We have followed here the
notation of Ref.~\cite{ferb97a}, which is slightly different from that
of the original work of Duplantier \cite{dupl99}, but more suitable for
a generalization to networks.  The star exponent can be written in the
following form:
\be
\eta (S) = - 2 \Delta(S) + f_1 \frac{\eta_\phi}{2} ,
\label{eq_etaS}
\ee
with $\Delta(S)$ the conformal scaling dimension and $\eta_\phi = 5/24$
the correlation length exponent \cite{ferb97a}.

Here we review briefly the main formulas leading to the exact value of
the exponent $\Delta(S)$ for an arbitray star copolymer $S$. Details of
the derivation can be found in the Refs. \cite{dupl99,dupl03}.  The main
idea is to map the star copolymer from the planar euclidean geometry onto
a two dimensional random lattice, i.e.  in the presence of {\it quantum
gravity}. 

Besides the bulk scaling dimension $\Delta(S)$ of a star $S$ we will also
consider star copolymers confined in a half-plane with the origin near
the boundary of the plane, which defines the surface scaling dimension
$\widetilde{\Delta}(S)$.  We use the notations $\Delta^{\rm qg} (S)$
and $\widetilde{\Delta}^{\rm qg} (S)$ for bulk and surface conformal
dimensions in the random lattice (here ${\rm qg}$ stands for quantum
gravity).  Given two walks $A$ and $B$, with a common origin we indicate
(as in Ref.~\cite{dupl99}) with the symbol $A \lor B$ a star configuration
where the two walks are allowed to overlap each other and with $A \land B$
a configuration where $A$ and $B$ are non-overlapping.

The main result of the theory is that in the fluctuating geometry 
the surface conformal dimensions for two mutually avoiding walks
are additive \cite{dupl99} i.e.:
\be
\widetilde\Delta^{\rm qg} (A \land B) = \widetilde\Delta^{\rm qg} (A) 
+ \widetilde\Delta^{\rm qg} (B) .
\label{eq_int}
\ee
For transparent walks in the plane, due to the trivial factorization
of their partition functions one has:
\be
\widetilde\Delta (A \lor B) = 
\widetilde\Delta (A)  + \widetilde\Delta (B) .
\label{eq_uni}
\ee
We point out to a sort of duality between Eqs. (\ref{eq_int}) and
(\ref{eq_uni}). Eq. (\ref{eq_int}) states that non-intersecting walks,
become transparent to each other when placed in a fluctuating geometry.
The Eqs. (\ref{eq_int}) and (\ref{eq_uni}) can be generalized to any
nested star copolymer structure (see Ref.~\cite{dupl99}).

Conformal invariance relates the bulk and surface scaling dimensions in
the two geometries as \cite{dupl99}:
\beqn
\Delta &=& U(\Delta^{\rm qg} ) ,
\label{map_bulk}\\
\widetilde\Delta &=& U(\widetilde\Delta^{\rm qg} ) ,
\label{map_surf}
\eeqn
where $U(x) = \frac{x}{3} (1 + 2x)$ and the inverse $U^{-1} (x) = 
\frac 1 4 (\sqrt{24 x +1} - 1)$.
The previous equations follow from the work of Knizhnik, Polyakov and
Zamolodchikov \cite{kniz88}, who established the existence of a relation
between the conformal dimensions of scaling operators in the plane and
those in the presence of gravity.
Finally the relation
\be
\widetilde\Delta^{\rm qg} = 2 \Delta^{\rm qg} + \frac 1 2 ,
\label{surf_bulk}
\ee
connects surface and bulk conformal dimensions in the fluctuating geometry
and can be derived from some factorization properties of star partition
functions under quantum gravity \cite{dupl98}.

For random and self-avoiding walks the conformal dimensions are
\cite{dupl99}:
\be
\widetilde\Delta_{\rm RW} = 1, \ \ \widetilde\Delta_{\rm RW}^{\rm qg} = 1, \ \ 
\Delta_{\rm RW} = \frac 1 8, \ \ \Delta_{\rm RW}^{\rm qg} = \frac 1 4 ,
\ee
\be
\widetilde\Delta_{\rm SAW} = \frac 5 8, \ \widetilde\Delta_{\rm SAW}^{\rm qg} 
= \frac 3 4, \ \Delta_{\rm SAW} = \frac 5 {96}, \
\Delta_{\rm SAW}^{\rm qg} = \frac 1 8 .
\ee

\begin{figure}[t]
\includegraphics[width=8cm]{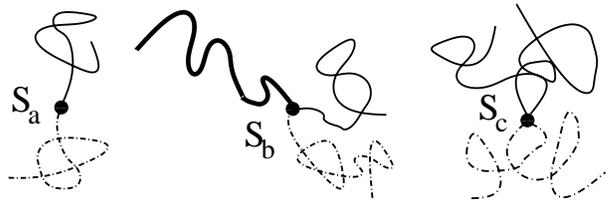}
\caption{
Star copolymers considered in this paper. Thin lines denote random
walks, while thick lines denote self-avoiding walks. Solid and dashed
thin lines are allowed to overlap with thin lines of the same species,
but not allowed to overlap thin lines of the other species (eg. overlap
between solid and dashed thin lines is not allowed).  The conformal
scaling dimensions for the examples in the figure are $\Delta (S_a) =
5/8$, $\Delta (S_b) = 39/32$ and $\Delta (S_c) = 35/24$ (see text).
}
\label{FIG04}
\end{figure}

As a practical example of the use of the above formulas we calculate
the conformal dimensions of the three star copolymers shown in
Fig.~\ref{FIG04}, which are the configurations relevant for the
model discussed in this paper.

Let us consider first the star copolymer composed by two mutually
avoiding random walks (i.e. $S_a$ of Fig.~\ref{FIG04}).  The equation
(\ref{eq_int}) states that the surface conformal dimension in the
fluctuating geometry for two mutually avoiding walks is additive, thus
for $S_a$ it is twice as large as that of a random walk
\be
\widetilde{\Delta}^{\rm qg} (S_a) = 2 \widetilde{\Delta}_{\rm RW}^{\rm qg} 
= 2 .
\ee
From Eq. (\ref{surf_bulk}) one finds therefore for the bulk dimension
in the fluctuating geometry: ${\Delta}^{\rm qg} (S_a) = 3/4$. 
The final step to obtain the bulk dimension in the plane is to use the
conformal mapping [Eq. (\ref{map_bulk})] which yields
\be
\Delta (S_a) = U (3/4) = \frac 5 8 .
\label{DeltaS_a}
\ee
The generalization to $k$ mutually avoiding random walks is given in
the Appendix \ref{sec:app1}. The above derivation of $\Delta (S_a)$
illustrates the general strategy of the calculation: the conformal
dimension is first calculated in the fluctuating geometry, where mutual
avoidance is easy to implement [see Eq. (\ref{eq_int})] and then obtained
for the planar geometry from Eqs. (\ref{map_bulk}) and (\ref{map_surf}).

For the star formed by two random walks and a self-avoiding walk all of
them avoiding each other ($S_b$ of Fig.~\ref{FIG04}), one proceeds along
the same lines as done for $S_a$. First, the additivity of surface scaling
dimensions in the fluctuating geometry [Eq. (\ref{eq_int})] implies that
$\widetilde{\Delta}^{\rm qg} (S_b) = 2 \widetilde{\Delta}_{\rm RW}^{\rm qg} +
\widetilde{\Delta}_{\rm SAW}^{\rm qg} = 11/4$. Therefore, for the bulk dimension
one finds [Eq. (\ref{surf_bulk})]: ${\Delta}^{\rm qg} (S_b) = 9/8$.
Finally, the conformal mapping [Eq. (\ref{map_bulk})] yields:
\be
\Delta (S_b) = U (9/8) = \frac{39}{32} .
\label{DeltaS_b}
\ee
For the star copolymer $S_c$ of Fig.~\ref{FIG04} the calculation is
slightly different. Both solid and dashed walks are transparent to each
other, therefore we first need to calculate the scaling dimension for the
sub-star composed either by solid or by dashed lines only. Two transparent
random walks (${\rm RW_1}$ and ${\rm RW_2}$) in the planar geometry have
additive surface conformal dimension [Eq. (\ref{eq_uni})] thus
\be
\widetilde{\Delta}({\rm RW_1} \lor {\rm RW_2}) = 2 \widetilde{\Delta}_{\rm RW} = 2 .
\ee
By inversion of Eq. (\ref{map_surf}), we obtain the corresponding surface
conformal dimension in the fluctuating geometry: $\widetilde{\Delta}^{\rm
qg} ({\rm RW_1} \lor {\rm RW_2}) = U^{-1} (2) = 3/2$. Now, as in the
fluctuating geometry the two scaling dimensions of the mutually avoiding
dashed and solid sub-stars are additive one finds from Eq. (\ref{eq_int}):
$\widetilde{\Delta}^{\rm qg} (S_c) =2 \widetilde{\Delta}^{\rm qg} ({\rm
RW_1} \lor {\rm RW_2}) = 3$. Eq. (\ref{surf_bulk}) yields for the bulk:
$\Delta^{\rm qg} (S_c) = 5/4$. The final step is the mapping back into
the planar geometry [Eq.(\ref{map_bulk})] which yields:
\be
\Delta (S_c) = U(5/4) = \frac{35}{24} .
\label{DeltaS_c}
\ee

\begin{figure}[t]
\includegraphics[width=7.5cm]{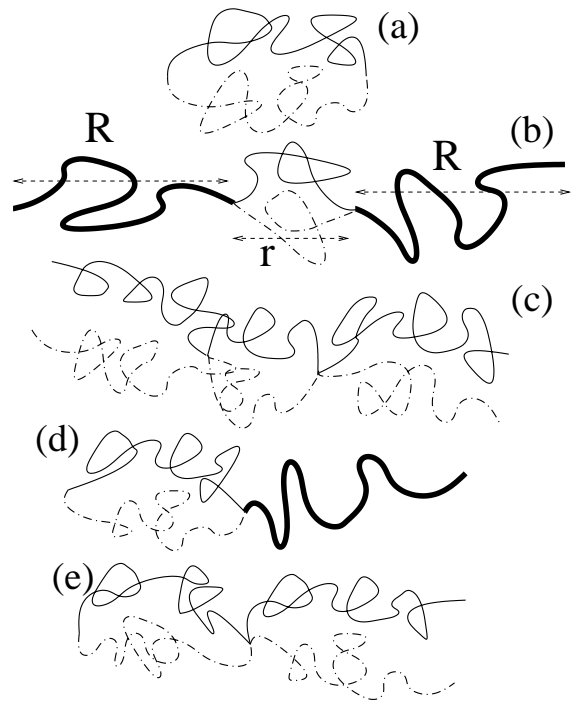}
\caption{Examples of loop configurations relevant for the model. (a)
Isolated loop formed by two mutually avoiding random walks, (b) loop
embedded between long SAWs ($R \gg r$), (c) Loop embedded between two
other long loops. At the edges the first loop formed at the origin of the
two walks can be attached to a long SAW (d) or to another long loop (e).
}
\label{FIG05}
\end{figure}

\subsection{Networks with arbitrary topology}

\begin{figure}[b]
\includegraphics[width=6.5cm]{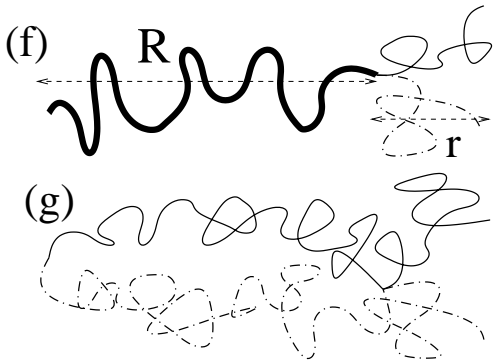}
\caption{
Configurations of end-segments linked to a long SAW (a) and to a loop formed
by two mutually avoiding random walks (b).
}
\label{FIG06}
\end{figure}

We consider now exponents associated to networks of arbitrary topology,
formed by mixtures of RWs and SAWs connected to each other in such a way
to form loops and dangling ends (examples are shown in Fig.~\ref{FIG05}
and \ref{FIG06}). The partition function for a network $G$ containing
$f_1$ SAWs, 
in the limit when all walks are critical takes the form
\cite{ferb97a}
\be
{Z} (G) \sim R^{\eta_G - f_1 \eta_2} ,
\label{ZG}
\ee
where the universal exponent $\eta_G$ depends on the topology of $G$
as follows \cite{ferb99}
\be
\eta_G = - d {\cal L} + \sum_{S} n_{S} \eta(S)
\label{etag}
\ee
where $d$ is the dimensionality of the system ($d=2$ in this paper),
${\cal L}$ the number of independent loops and the sum is extended to
all constituent vertices forming the network, each contributing for
a factor $\eta(S)$ (the star exponent defined by Eqs. (\ref{eq_etaS})
and (\ref{star_disc}), and $n_S$ is the degeneracy of the vertex $S$.
Using the results of the preceding section and recalling that for a
vertex $S$ with $f_1$ outgoing SAWs one has $\eta(S) = - 2 \Delta(S)+
f_1 \eta_\phi/2$ [Eq. (\ref{eq_etaS})] we can now calculate the network
exponents for the examples in Fig.~\ref{FIG05} and \ref{FIG06}, which
are those relevant for the unbinding transition considered in this paper.

For the isolated loop of Fig.~\ref{FIG05}(a) one has ${\cal L}=1$
and there are two vertices $S_a$, as defined in Fig.~\ref{FIG04}.
As there are no SAWs $f_1=0$ in Eqs. (\ref{ZG}) and (\ref{eq_etaS}) 
therefore Eqs. (\ref{etag}) and (\ref{DeltaS_a}) imply that
\be
\eta_G = -2 - 2 \eta (S_a) = -2 - 4 \Delta (S_a) = -2 - \frac{5}{2}
\ee
Usually we are interested in the scaling as function of the loop length,
and not of its radius of gyration, therefore the partition function is:
\be
{Z}_{\rm loop} \sim l^{\nu \eta_G} \sim l^{-c_0}
\ee
with $\nu=1/2$ for random walks and where we have introduced the
entropic exponent for an isolated loop, in analogy as what discussed
in the Introduction. We have $c_0 = 2 + 1/4 = 2.25$.  Differently from
the case of self-avoiding loops (see Introduction) in the present model
already at the level of an isolated loop $c_0 > 2$, implying that the
first order character of the transition is rather strong, as remarked
in Ref.~\cite{baie02}.

Equation (\ref{ZG}) can be generalized to the case where the network is
formed by walks of different sizes, say $R$ and $r$. In this case the
network partition function becomes:
\be
{Z} (G) \sim R^{\eta_G - f_1 \eta_2} f\left( \frac r R \right)
\label{part_copolym}
\ee
with $f$ a scaling function. In all the other cases
(Fig.~\ref{FIG05}(b-e)) we will consider the limit in which $r$, the
size of the loop, is much smaller than $R$ the size of the walks or
loops attached to it.

We consider now the network of Fig.~\ref{FIG05}(b) which contains one loop
${\cal L} = 1$ and two vertices $S_b$ (as in Fig.~\ref{FIG04}) therefore:
\be
\eta_G -2 \eta_2 = -2 + 2 \left[- 2 \Delta (S_b) + 
\frac{\eta_\phi}{2} - \eta_2 \right]
\ee
In the limit $r \ll R$ one should recover the partition function of a
single SAW ($\sim R^{- \eta_2}$), which implies that the scaling function
of Eq. (\ref{part_copolym}) behaves as
\be 
f(x) \sim x^{-2 - 4 \Delta (S_b) + \eta_\phi - \eta_2}  \ \ \ \ \ \ 
{\rm for} \ \ x \to 0
\ee
In this limit the partition function of the network becomes:
\be
Z(G) \sim R^{- \eta_2}  r^{-2 - 4 \Delta (S_b) + \eta_\phi - \eta_2}
\ee
and thus factorizes as $Z(G) \sim Z_{\rm SAW} Z_{\rm loop}$, i.e. in
a contribution from a long SAW and of a loop. The latter expressed
in terms of its total length $l$ reads:
\be
{Z}_{\rm loop} \sim l^{- \nu [2 + 4 \Delta (S_b) - \eta_\phi + \eta_2]} 
\sim l^{- c_{\rm ss}}
\label{cll}
\ee
Using the numerical values of the exponents given in the preceding
section we find: $c_{\rm ss} = 3 + 5/32$ \cite{correct}. Note that the
derivation of Eq. (\ref{cll}) is similar to that for the unbinding of
SAWs reported in Ref.~\cite{kafr00}.

The next example is the configuration of Fig.~\ref{FIG05}(c), i.e. a
loop confined between two other long loops. As the calculation of its
partition function follows closely that of a loop confined between two
SAWs we report here the final result:
\be
{Z}_{\rm loop} \sim l^{- \nu [2 + 2 \Delta(S_c)]} \sim l^{- c_{\rm ll}} ,
\ee
and thus from Eq. (\ref{DeltaS_c}) we find: $c_{\rm ll} = 2 + 11/24$.
Notice that $c_{\rm ll} < c_{\rm ss}$, contrary to what happens for
self-avoiding walks \cite{kafr00}.  This follows from the random
walks character for isolated strands: As solid and dashed lines in
Fig.~\ref{FIG05}(c) are allowed to overlap themselves a loop bounded
by two loops is "less localized" (smaller $c$) than a loop bounded by
two SAWs.

We will also be interested in the statistical properties of the first
loop formed at the common origin of the two walks, therefore we consider
also the case of a loop bound to the rest of the polymer only on one edge
(see Fig.~\ref{FIG05}(d-e)).
Also in this case we only report the final results as the calculation 
follows the example above. For the partition function of a loop bound 
to a long segment (Fig.~\ref{FIG05}(d)) we find
\be
{Z}_{\rm loop} \sim l^{-\nu[2 +2\Delta (S_a)+2\Delta (S_b)-\eta_\phi/2]} 
\sim l^{-c_{\rm s}}
\ee
with $c_{\rm s} = 2 + 19/24$ while for a loop attached to a long
loop (Fig.~\ref{FIG05}(e)) we find:
\be
{Z}_{\rm loop} \sim l^{-\nu [2+2\Delta (S_c)]} \sim l^{- c_{\rm l}}
\ee
with $c_{\rm l} = 2 + 11/24$. Notice that, curiously, $c_{\rm l}= c_{\rm
ll}$ a relation which is not valid only for this particular model, but
it is quite general for all polymer unbinding transitions, wether the
constituents strands are SAWs or mutually avoiding RWs.

Finally besides loops it is also interesting to consider end-segments
distributions, i.e the length of the single strands at the free
end of the polymer (see Fig.~\ref{FIG06}), as done for the SAWs in
Ref.~\cite{kafr02}.
Again, as the calculation is very similar to those reported in this
section we only give the final results. For end segments each of length
$n$ bounded to a SAW of size $R$, in the limit $R \gg r$, with $r$
the size of the end-segments ($r \sim n^{1/2}$) we find:
\be
Z_{\rm end} \sim n^{-\nu [2 \Delta (S_b) - \eta_\phi/2]} 
\sim n^{-\gamma_s}
\ee
where we have introduced a new exponent equal to $\gamma_s = 7/6$.

Similarly for the configuration of Fig.~\ref{FIG06}(g) we find
\be
{Z}_{\rm end} \sim  n^{-\nu[2 \Delta(S_c) - 2 \Delta(S_a)]} 
\sim n^{-\gamma_l}
\ee
with $\gamma_l = 5/6$. Notice that analogously as what we have found for the
loops, also the end-segments bound to a long loop are less localized than
those bound to a SAW ($\gamma_l < \gamma_s$).

\begin{table}[t]
\caption{Summary of the exact loop and end-segments entropic exponents for
the configurations of Fig.~\ref{FIG05} and \ref{FIG06}.}
\begin{ruledtabular}
\begin{tabular}{cc|cc}
 & Loops  &  & End segments \\
\hline
       $c_0$ &   $2+1/4 \ (\approx 2.25)$ &       $\gamma_0$ & $5/8 \ (\approx 0.62)$\\
$c_{\rm ss}$ & $3+5/32  \ (\approx 3.16)$ & $\gamma_{\rm s}$ & $7/6 \ (\approx 1.17)$\\
$c_{\rm ll}$ & $2+11/24 \ (\approx 2.46)$ & $\gamma_{\rm l}$ & $5/6 \ (\approx 0.83)$\\
 $c_{\rm s}$ & $2+19/24 \ (\approx 2.79)$ &  &   \\
 $c_{\rm l}$ & $2+11/24 \ (\approx 2.46)$ &  & 
\end{tabular}
\end{ruledtabular}
\label{tableI}
\end{table}

The results obtained in this Section are summarized in Table \ref{tableI},
to which we have also added $\gamma_0$ the exponent associated to
isolated end-segments. This exponent is associated to the configuration
of Fig.~\ref{FIG04}(a), therefore: $\gamma_0 = 2 \nu \Delta(S_a) = 5/8$.

\begin{figure}[b]
\includegraphics[width=8cm]{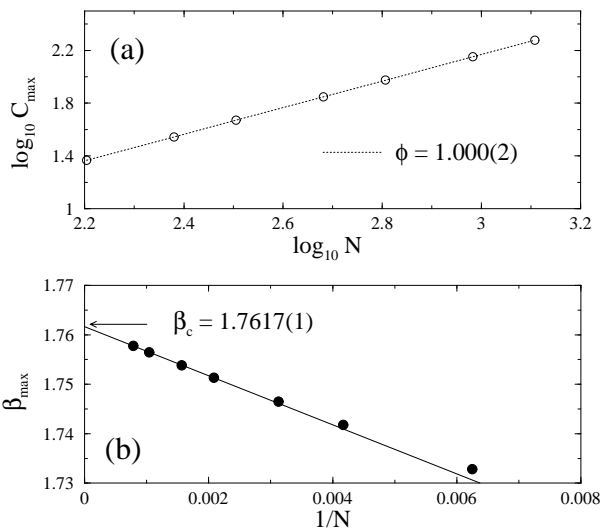}
\caption{(a) Plot of the specific heat peaks heights $C_{\rm max}$ as a
function of the chain length $N$ on a log-log scale. From a linear fit
and by Eq. (\ref{cmax}) we find as estimate of the crossover exponent
$\phi = 1.000(2)$. (b) Plot of the inverse temperature position of the
peaks as function of $1/N$.}
\label{FIG07}
\end{figure}

\section{Numerical results}
\label{sec:numerics}

Polymer configurations in which each strand is of length $N=1280$
were generated by the Pruned Enriched Rosenbluth Method (PERM) which
is described in Ref.~\cite{gras97}. We performed first three runs at
fixed temperatures around the critical point and used the multihystogram
method~\cite{ferr88} 
to interpolate results to arbitrary temperatures in the transition
region. A precise estimate of the transition point was obtained by the
analysis of the specific heat maximum per unit of length $C_{\rm max}
(N)$, which is expected to scale as function of the chains length $N$
as \cite{vand98}
\be
C_{\rm max} (N) \sim N^{2 \phi -1}
\label{cmax}
\ee
a relation which defines the crossover exponent $\phi$. Figure
\ref{FIG07}(a) shows a plot of the peak heights $C_{\rm max} (N)$
as a function of the chain length $N$ on a log-log scale. A linear
fit of the data yields $\phi = 1.000(2)$ in excellent agreement with a
first order transition, for which one expects $\phi=1$ \cite{vand98}.
The sharp determination of $\phi$ is a signature of a rather strong first
order character of the transition. It is interesting to point out that in
the case of unbinding of self-avoiding walks, although the transitions
is known to be of first order type, it is difficult to extrapolate an
exponent $\phi$ which is consistent with $\phi = 1$ \cite{caus00,carl02}
(at least in three dimensions).

The peak position is expected to scale as 
\be
\beta_{\rm max} (N) \sim \beta_c + \frac{A}{N^{\phi}}
\ee
with $A$ a constant. The first order character ($\phi=1$) is confirmed by
the scaling of the peak positions, as illustrated in Fig.~\ref{FIG07}(b)
which shows a plot of $\beta_{\rm max} (N)$ vs. $1/N$ ($\beta$ is the
inverse temperature).  We performed a series of iterated linear fits of
$\beta_{\rm max} (N)$ vs. $1/N$ and obtained the following estimate of
the transition point $\beta_c = 1.7617(1)$.

\begin{figure}[t]
\includegraphics[width=8cm]{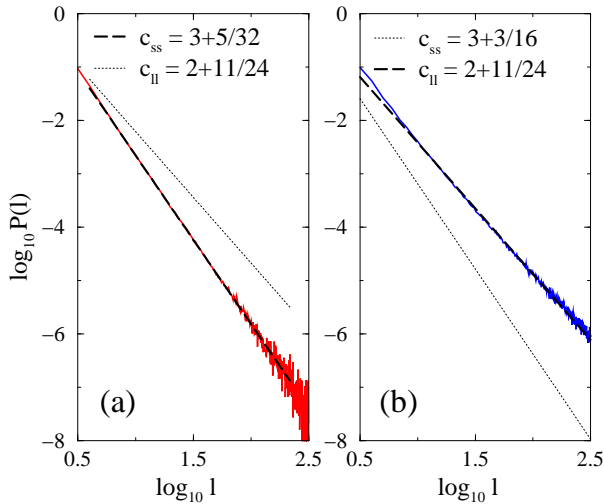}
\caption{Probability distribution of loop lengths at the estimated
transition point $\beta= \beta_c = 1.7617$ (a) and in the high temperature
phase $\beta = 1.6$ (b).  As references we plot the exact exponents for
a loop embedded between two segments $c_{\rm ss} = 3 + 5/32$ and a loop
embedded between two long loops $c_{\rm ll} = 2 + 11/24$, calculated in
the preceding section.}
\label{FIG08}
\end{figure}

\subsection{Behavior at the transition point $\beta = \beta_c$}

We focus first on the behavior at the transition point $\beta = \beta_c
= 1.7617$. The probability distribution of finding a loop of length $l$
is expected to decay as a function of $l$ as \cite{carl02}
\be
P(l) \sim l^{-c} ,
\ee
from which the exponent $c$ can be calculated. Figure \ref{FIG08}(a)
shows a log-log plot of $P(l)$ versus $l$ for $N=1280$ at $\beta_c$. As
a comparison we plot, as straight lines, the slopes corresponding to the
analytical estimates of the exponents $c_{\rm ss}$ and $c_{\rm ll}$,
for a loop embedded between two segments and two loops, respectively,
calculated in the preceding section. Notice the excellent agreement of the
numerical results with the decay exponent $c_{\rm ss}$, which indicates
that considering each loop as simply bounded by pure SAWs approximates
extremely well the polymer configuration at the transition point.
As $P(l)$ decays rather fast in $l$ ($\sim l^{-c}$ with $c \approx 3.2$),
rather long runs are needed in order to obtain an accurate statistics.

\begin{figure}[t]
\includegraphics[width=8cm]{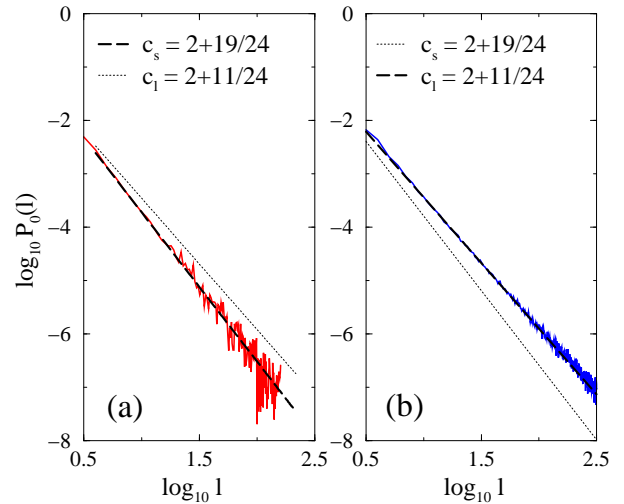}
\caption{Probability distribution of the first loop as a function of its
length in a log-log scale at the estimated transition point $\beta =
\beta_c = 1.7617$ (a) and in the high temperature phase $\beta = 1.6$
(b). The slopes corresponding to the analytical estimates for the
exponents $c_s$ and $c_l$ for a loop bounded by a SAW segment and by
another long loops are also shown.}
\label{FIG09}
\end{figure}

We have also analyzed the probability distribution of the first loop $P_0
(l)$, which is formed at the common origin of the two strands. Here we
have considered only loops originated at the very edge for which the
first monomer is unbound ($\vec{r}_1 (1) \neq \vec{r}_2 (1)$). A plot
of $\log P_0 (l)$ as a function of $\log l$ at $\beta_c$ is shown in
Fig.~\ref{FIG09}(a).  The statistics is poorer compared to the total loop
distribution in Fig.~\ref{FIG08}(a), as in most of the configurations the
first monomer is bound, therefore loops at the common edge of the two
strands are quite rare. Despite that, the agreement with the expected
exponent $c_s = 2+19/24$ is very good.

Finally we considered the end-segments distribution which is shown in
Fig.~\ref{FIG10}(a). Here $P_{\rm e}(n)$ is the probability of having
a configuration in which the last $n-1$ monomers are unbound  while
$\vec{r}_1 (n) = \vec{r}_2 (n)$.  Once again at $\beta_c$ we note a
good agreement with a decay $P_{\rm e}(n) \sim n^{-\gamma_s}$ with an
exponent $\gamma_s = 7/6$.

\subsection{Behavior in the high temperature phase $\beta < \beta_c$}

We have repeated the same type of calculation of loops and end-segments
statistics also in the high temperature region $\beta < \beta_c$. We
expect a power-law distribution of loop lengths also at high temperatures
\cite{carl02}.

Figure \ref{FIG08}(b) shows a plot of the loop probability distribution
at $\beta=1.6$ for $N=1280$. Indeed there is a clear power-law decay
governed by an exponent which is in excellent agreement with $c_{\rm
ll} = 2.46$ as calculated in the preceding section. Typically at high
temperatures a loop is more likely to be bound by neighboring loops,
rather than double stranded segments, as contacts between the two strands
are rare, which explains the observed exponent.

\begin{figure}[t]
\includegraphics[width=8cm]{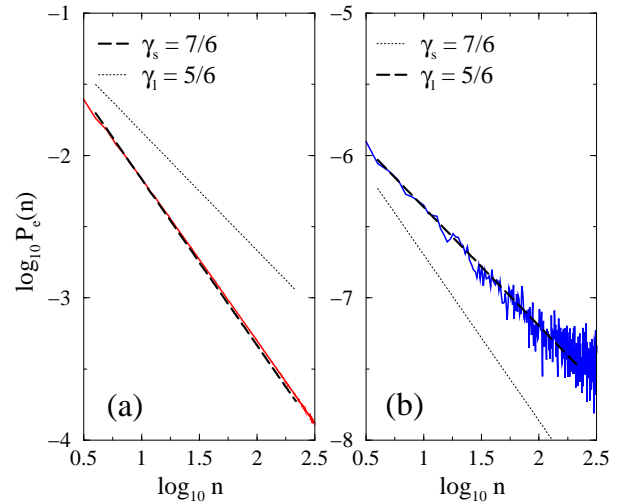}
\caption{Probability distribution for the end segments at the estimated
transition point $\beta = \beta_c = 1.7617$ (a) and in the high
temperature region $\beta = 1.6$ (b). The analytical estimates $\gamma_s$
and $\gamma_l$ for the slopes for end segments attached to a SAW and to
a loop respectively, are also shown.}
\label{FIG10}
\end{figure}

Analogously, also for the first loop length distribution $P_0 (l)$,
we find a decay exponent in very good agreement with $c_l$ (see
Fig.~\ref{FIG09}(b)). Notice that, differently from the $\beta=\beta_c$
case, here the distribution is rather smooth, as, in the high temperature
region, it is more likely to find configuration where a loop forms at
the origin.

An analogous very good agreement with the exponent $\gamma_l$ has also
been found for the decay of the end segments distribution $P_{\rm e}
(n)$, as illustrated in Fig.~\ref{FIG10}(b).

\section{Discussion}
\label{sec:discussion}

In this paper we have studied the unbinding transition for a
two-dimensional lattice polymer composed of two strands which are random
walks mutually avoiding each other. Considering all the polymer unbinding
models studied so far in the literature in two and three dimensions, the
present model is that with the strongest first order transition (higher
loop exponent $c\approx 3.2$). The sharp first order behavior can be
inferred from the determination of the crossover exponent $\phi$, which
is in excellent agreement with the first order value $\phi=1$, already
clearly observed for rather short chains ($N \approx 100$). This can be
compared with, for instance, the unbinding for SAWs in three dimensions
for which $c \approx 2.1$ (weak first order, just above the threshold $c
=2$) and numerical $\phi \sim 0.9$, in itself not fully consistent with
a first order transition \cite{caus00}.

We showed that numerical estimates of entropic exponents for loops
and end segments are in excellent agreement with analytical results,
obtained from a recent theory based on mapping of star copolymers into a
fluctuating geometry \cite{dupl99}. Such good agreement is not surprising
in the high temperature phase where the loops are most likely bound
by other loops as contacts between the strands are very rare.  At the
transition point one expects that typical configurations are composed
by a double stranded polymer "dressed" with loops of all sizes. Notice
however that as the loop exponent is rather large ($c \approx 3.2$),
the statistical weight of long loops is suppressed. In particular $c>3$
implies that the two first moments $\langle l \rangle$ and $\langle
l^2 \rangle$ are finite. In this case, as loops are typically very
small (see also Fig.~\ref{FIG02}), therefore neglecting totally their
effect is still a very good approximation.  Thus the strong first order
unbinding represents an ideal case where polymer networks calculations
work extremely well.

\begin{figure}[b]
\includegraphics[width=7cm]{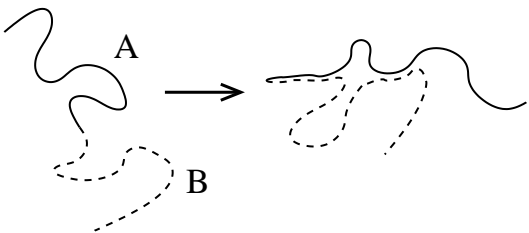}
\caption{Zipping transition for a two dimensional diblock copolymer. 
Both strands A and B are mutually and self-avoiding. Loops in the zipped
phase may have different lengths in the two strands.
}
\label{FIG11}
\end{figure}

As a counterexample we mention the case of a two dimensional unbinding
transition studied recently in diblock copolymers \cite{baie01}. A diblock
copolymer is composed by two homogeneous branches of A and B monomers
joined at a common origin. In the model of Ref.~\cite{baie01} A and B are
both self- and mutually avoiding.  An attractive interaction between
A and B induces a "zipping" transition by lowering the temperature
where the two strands are bound (see Fig.~\ref{FIG11}). Differently
from the model studied here and in models of DNA denaturation
\cite{pola66,kafr00,caus00,carl02} any "monomer" in A can bind to any
"monomer" in B, therefore loops of total length $l$ may have different
lengths along the two strands, i.e. $l = l_A + l_B$ with $l_A \neq l_B$.
Due to this freedom the loop entropic exponents for the diblock copolymer
are given by:
\beqn
\label{db_ss}
c_{\rm ss} = c_{\rm ss}^{\rm (SAW)} - 1 \approx 1.42 \\
\label{db_ll}
c_{\rm ll} = c_{\rm ll}^{\rm (SAW)} - 1 \approx 1.64 
\eeqn
where we have used the two dimensional SAW exponents \cite{kafr00}:
$c_{\rm ss}^{\rm (SAW)} = 2.42$ and $c_{\rm ll}^{\rm (SAW)} = 2.64$.
Numerical results \cite{baie01} shows that the zipping transition is
continuous and the specific heat exponent is in excellent agreement with
$\phi = 9/16$, conjectured to be an exact value. From scaling arguments
for a continuous transition \cite{carl02} one has $c = 1 + \phi = 25/16
\approx 1.56$, which is in between the values of Eqs. (\ref{db_ss}) and
(\ref{db_ll}), clearly distinct from both. At the zipping transition
therefore the polymer network theory does not reproduce the numerical
value for the loops entropic exponent as accurately as for the model
studied here. Notice that however, also in this case the numerically
determined $c$ satisfies the relation $c_{\rm ss} < c < c_{\rm ll}$, 
as expected.

\acknowledgements

We are grateful to C.~von~Ferber, Yu.~Holovatch, E.~Orlandini and
A.~L.~Stella for fruitful discussions.
The work was supported by INFM-PAIS02.

\appendix

\section{Star copolymer made of $k$ mutually avoiding random walks}
\label{sec:app1}

We generalize here the calculation leading to Eq. (\ref{DeltaS_a})
to the case of consider a star copolymer $S_k$ made of $k$ mutually
avoiding random walks.  In this case the surface conformal dimension
in the fluctuating geometry is $k$-times that of a single random
walk [Eq. (\ref{eq_int})]: $\widetilde{\Delta}^{\rm qg} (S_k) =k$.
Using Eqs. (\ref{surf_bulk}) and (\ref{map_bulk}) we find:
\be
\Delta (S_k) = \frac{4 k^2 - 1}{24}
\ee
which correctly reproduces $\Delta (S_a)$ of Eq. (\ref{DeltaS_a})
for $k=2$. In the case of four mutually avoiding walks
\be
\Delta (S_4) = \frac{21}{8}
\ee
Notice that the vertex $S_c$ of Fig.~\ref{FIG04} is also formed by four
random walks, however not all mutually avoiding and its bulk conformal
dimension is $\Delta (S_c) = 35/24 < \Delta (S_4)$.  $\Delta (S_c)$
is smaller as more configurations are available for the star copolymer
when partial overlapping between walks is allowed, as in $S_c$. Had
we put no restrictions on the order of the loops in the construction
of the model both vertices $S_c$ and $S_4$ would have been generated.
It should be emphasized that for two dimensional star copolymers the
order of the constituing walks does matter.


\begin{thebibliography}{24}
\bibitem[{cor()}]{correct}
\bibinfo{howpublished}{This corrects an error of Ref. \cite{baie02b} where the
  value 3+7/16 was reported}.
\end{thebibliography}
\end{document}